\documentclass[
 reprint, 
 superscriptaddress,
 amsmath,amssymb,
 aps, 
prc,
]{revtex4-2}

\usepackage{calc}
\usepackage[caption=false]{subfig}
\usepackage{graphicx}
\usepackage{dcolumn}
\usepackage{bm}
\usepackage{xcolor} 
\usepackage{comment} 
\usepackage{hyperref}
\usepackage{upgreek}
\usepackage[normalem]{ulem} 
\usepackage{pgf}

\newcommand{\gram}	{\ensuremath{\,\text{g}}}

\newcommand{\microm}	{\ensuremath{\,\upmu\text{m}}}
\newcommand{\mm}	{\ensuremath{\,\text{mm}}}
\newcommand{\cm}	{\ensuremath{\,\text{cm}}}

\newcommand{\keV}	{\ensuremath{\,\text{keV}}}

\newcommand{\ns}	{\ensuremath{\,\text{ns}}}
\newcommand{\micros}	{\ensuremath{\,\upmu\text{s}}}

\newcommand{\seconds}	{\ensuremath{\,\text{s}}}
\newcommand{\minute}	{\ensuremath{\,\text{min}}}
\newcommand{\hour}	{\ensuremath{\,\text{h}}}
\newcommand{\days}	{\ensuremath{\,\text{d}}}

\newcommand{\kHz}	{\ensuremath{\,\text{kHz}}}

\newcommand{\ppt}	{\ensuremath{\,\text{ppt}}}

\newcommand{\gammaray}	{\ensuremath{\upgamma\text{-ray}}}

\newcommand{\neutron}	{\ensuremath{\text{n}}}
\let\oldalpha\alpha
\renewcommand{\alpha}	{\ensuremath{\oldalpha}}

\newcommand{\betadecay}	{\ensuremath{\upbeta\text{-decay}}}

\newcommand{\gammagamma}{\ensuremath{\upgamma\upgamma}}
\newcommand{\gammagammacounting}{\ensuremath{\gammagamma\text{ coincidence-counting}}}
\newcommand{\gammagammacountingnohyphen}{\ensuremath{\gammagamma\text{ coincidence counting}}}
\newcommand{\gammacounting}{\ensuremath{\upgamma\text{ counting}}}
\newcommand{\flyash}{\ensuremath{\text{fly ash}}}

\newcommand{\natwentyfour}	{\ensuremath{^{24}\text{Na}}}
\newcommand{\altwentyseven}	{\ensuremath{^{27}\text{Al}}}

\newcommand{\scfortysix}	{\ensuremath{^{46}\text{Sc}}}
\newcommand{\mnfiftyfour}	{\ensuremath{^{54}\text{Mn}}}
\newcommand{\cofiftyseven}	{\ensuremath{^{57}\text{Co}}}
\newcommand{\cosixty}		{\ensuremath{^{60}\text{Co}}}
\newcommand{\znsixtyfive}	{\ensuremath{^{65}\text{Zn}}}
\newcommand{\yeightyeight}	{\ensuremath{^{88}\text{Y}}}
\newcommand{\moninetynine}	{\ensuremath{^{99}\text{Mo}}}
\newcommand{\cdoneonine}	{\ensuremath{^{109}\text{Cd}}}
\newcommand{\snonethirteen}	{\ensuremath{^{113}\text{Sn}}}
\newcommand{\csonethirtyfour}	{\ensuremath{^{134}\text{Cs}}}
\newcommand{\csonethirtyseven}	{\ensuremath{^{137}\text{Cs}}}

\newcommand{\ceonethirtynine}	{\ensuremath{^{139}\text{Ce}}}
\newcommand{\hfoneeightyone}    {\ensuremath{^{181}\text{Hf}}}
\newcommand{\thtwothirtytwo}	{\ensuremath{^{232}\text{Th}}}

\newcommand{\utwothirtyeight}	{\ensuremath{^{238}\text{U}}}
\newcommand{\nptwothirtynine}	{\ensuremath{^{239}\text{Np}}}
\newcommand{\putwothirtynine}   {\ensuremath{^{239}\text{Pu}}}
\newcommand{\amtwofortyone}	{\ensuremath{^{241}\text{Am}}}

\begin{document}

\preprint{arXiv:2508.04232}

\title{\textbf{Enhanced sensitivity to trace $^{238}$U impurity of sapphire via coincidence neutron activation analysis}}

\author{D.~Chernyak}
\email{Corresponding author: d.n.chernyak@gmail.com}
\altaffiliation[Now at ]{Research Center for Neutrino Science, Tohoku University, Sendai 980-8578, Japan}
\affiliation{Department of Physics and Astronomy, University of Alabama, Tuscaloosa, AL 35487, USA}
\author{I.J.~Arnquist}
\affiliation{Pacific Northwest National Laboratory, Richland, WA 99352, USA}
\author{T.~Daniels}
\affiliation{Department of Physics and Physical Oceanography, University of North Carolina Wilmington, Wilmington, NC 28403, USA}
\author{S.W.~Finch}
\affiliation{Department of Physics, Duke University and Triangle Universities Nuclear Laboratory, Durham, NC 27708-0308, USA}
\author{L.~Hissong}
\affiliation{Department of Physics and Astronomy, University of Kentucky, Lexington, KY 40506, USA}
\author{M.~Hughes}
\affiliation{Department of Physics and Astronomy, University of Alabama, Tuscaloosa, AL 35487, USA}
\author{R.~MacLellan}
\affiliation{Department of Physics and Astronomy, University of Kentucky, Lexington, KY 40506, USA}
\author{A.~Piepke}
\affiliation{Department of Physics and Astronomy, University of Alabama, Tuscaloosa, AL 35487, USA}
\author{A.~Pocar}
\affiliation{Amherst Center for Fundamental Interactions and Physics Department, University of Massachusetts Amherst, MA 01003, USA}
\author{R.~Roshong}
\affiliation{Department of Physics and Astronomy, University of Kentucky, Lexington, KY 40506, USA}
\author{R.~Saldanha}
\affiliation{Pacific Northwest National Laboratory, Richland, WA 99352, USA}
\author{R.H.M.~Tsang}
\altaffiliation[Now at ]{Canon Medical Research USA, Inc.}
\affiliation{Department of Physics and Astronomy, University of Alabama, Tuscaloosa, AL 35487, USA}

\date{December 2, 2025}

\begin{abstract}
Sapphire has mechanical and electrical properties that are advantageous for the construction of internal components of radiation detectors such as time projection chambers and bolometers. However, it has proved difficult to assess its $\rm ^{232}Th$ and $\rm ^{238}U$ content down to the picogram per gram level. This work reports an experimental verification of a computational study that demonstrates \gammagammacountingnohyphen, coupled with neutron activation analysis (NAA), can reach ppt sensitivities. Combining results from \gammagammacountingnohyphen\ with those of earlier single-$\upgamma$ counting based NAA shows that a sample of Saint Gobain sapphire has $\rm ^{232}Th$ and $\rm ^{238}U$ concentrations of $<0.26$\ppt\ and $<2.3$\ ppt, respectively; the best constraints on the radiopurity of sapphire. 
\end{abstract}

\maketitle


\section{\label{sec:intro}Introduction}
Sapphire is a material of interest for a number of rare-event experiments. The low atomic mass of aluminum and oxygen make it an attractive target for detecting low energy recoils from dark matter or coherent neutrino nucleus scattering \cite{verma2023low}. The polar crystal structure can couple to dark matter to produce acoustic and optical phonons \cite{griffin2018directional}. 
Low-threshold phonon and scintillation detection has been demonstrated with sapphire targets at cryogenic temperatures in the ROSEBUD and CRESST experiments~\cite{rosebud_1999, cresst2024first} and it is also a candidate material for the SPICE experiment \cite{billard2024transition}. Such applications of sapphire require low-radioactivity levels. Radiopure sapphire has also been proposed as an interposer substrate to reduce the amount of ionizing radiation to which superconducting quantum devices are exposed \cite{quantum_devices_2024}. 

The development of the new analysis technique, discussed in this paper, was motivated by the low-radioactivity material requirements of large time projection chambers (TPCs) being used in ongoing~\cite{darkside-50,deap_2020,LZ_2023,next_2023,pandax-4t_2023,xenon_nt_2023} and planned rare event searches~\cite{darkside-20k,Darwin_2016,next_ts_2023}, with specific application to the development of the next-generation double beta decay search nEXO~\cite{nEXO:2022nam}. Filled with different noble elements, their ability to reconstruct event energies, tracks, and locations gives them a powerful background recognition capability, and the use of a fluid target offers scalability to ton-size devices.
Achieving a uniform electric drift field across the TPC requires the installation of internal electrode structures, which need to be mechanically supported and electrically isolated from other components. 
For low-rate, low-energy experiments, the support material, as all other components, needs to be as radiopure as possible. 

Sapphire ($\rm Al_2O_3)$, fused silica ($\rm SiO_2$), polytetrafluoroethylene (PTFE; $\rm (C_2F_4)_n$), and polyetherimide (PEI; $\rm (C_{37}H_{24}O_6N_2)_n$) are candidate electrically-insulating materials for the mechanical support of large electrodes and other high voltage components in low-background TPCs. Among them, sapphire has the most favorable mechanical properties, exhibiting the largest Young's modulus and the highest tensile and compressive strengths. 
However, these advantages are offset by the difficulty in precisely determining the trace levels of $\rm ^{232}Th$ and $\rm ^{238}U$ contained in sapphire.

At concentrations of 1\ppt, a benchmark value for many rare event searches, decay counting using \gammaray\ spectroscopy has to determine specific activities of $0.35\,\text{decays}/(\text{kg}\cdot\text{d})$ and $1.1\,\text{decays}/(\text{kg}\cdot\text{d})$ for Th and U,
requiring large, costly samples. 
Inductively coupled plasma mass spectrometry (ICP-MS) requires the analyte to be dissolved in mineral acids. Sapphire is largely insoluble in such acids. 
The assay sensitivities achievable by neutron activation analysis (NAA), using a high-flux research reactor as neutron source, are limited by large $\rm ^{24}Na$ side activities, created by the parasitic co-activation of the sapphire itself via the fast neutron reaction $\altwentyseven(\neutron, \upalpha)\natwentyfour$. 
The co-activation of various chemical impurities leads to sizable side activities limiting the sensitivity for $\rm ^{232}Th$ and $\rm ^{238}U$.
The nEXO Collaboration achieved sub-ppt sensitivity for $\rm ^{232}Th$ but only 11 ppt for $\rm ^{238}U$~\cite{tsang_2021}. 

\begin{figure}[bt]
\includegraphics[width=\columnwidth]{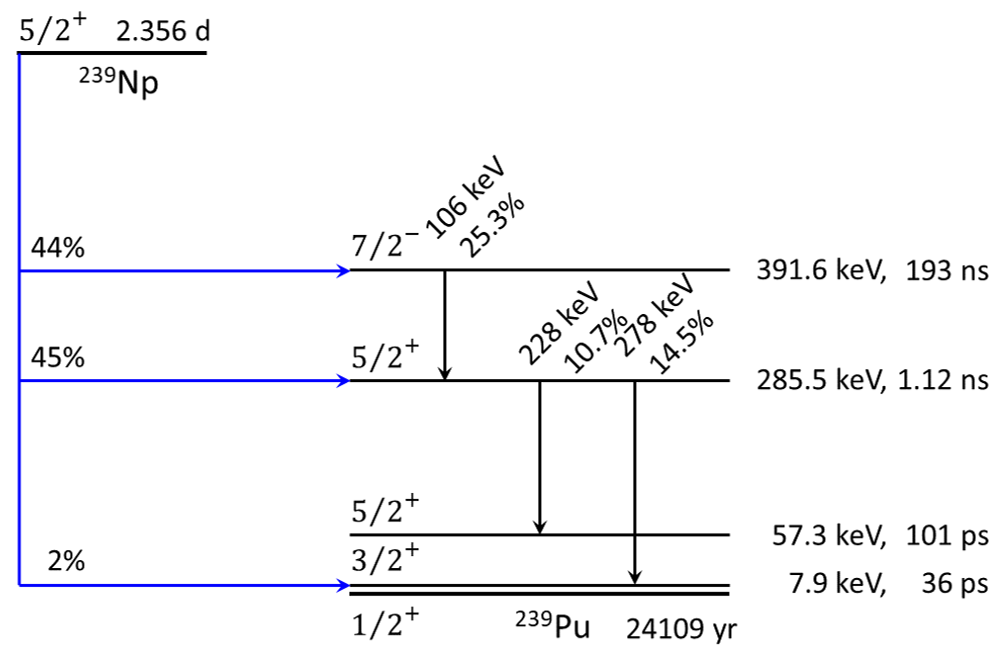}
\caption{\label{fig:239np_partial_level_scheme} 
Partial level scheme of \putwothirtynine. Energies, transition strengths, spin-parity assignments and half-lives are taken from Ref.~\cite{Browne2014}. The transitions used in this study are indicated.}
\end{figure}

A computational study showed that the limitation to NAA of sapphire can be overcome by using, Ge-detector based, \gammagamma\ coincidence spectroscopy~\cite{tsang_2021}. 
This technique exploits the correlated emission of $\gamma$ radiation in the form of two cascades within the $\rm ^{239}Pu$ level scheme, shown in Fig.~\ref{fig:239np_partial_level_scheme}. 
The partial level scheme indicates the coincident transitions with 106 keV-228 keV and 106 keV-278 keV used in this work.
The decay of the $\rm ^{232}Th$ activation product $\rm ^{233}Pa$ does not offer a convenient $\gamma$ cascade.
The application of the \gammagamma\ coincidence technique to NAA has been reported before~\cite{tomlin_2008,zeisler_2017,drescher_2018}; however, not focused on the determination of trace amounts of uranium in the presence of large side activities.

The work presented here is an experimental demonstration of the method described in Ref.~\cite{tsang_2021}. It is suitable for boosting the NAA assay sensitivity for $\rm ^{238}U$ for a broad range of materials, where large sample-related backgrounds are the limiting factor. 
This paper discusses the measurement approach, data analysis technique, and systematic quality checks. 
As shown in section~\ref{data_analysis},
its application allows us to derive the most stringent $\rm ^{238}U$ assay result for sapphire. 

The chosen method is sensitive to the long-lived head of the $\rm ^{238}U$-decay series. It does not allow inferring the concentration of $\rm ^{226}Ra$ or $\rm ^{210}Pb$ and, with it, the possible breakage of secular chain equilibrium.

\section{\label{sec:setup}Measurement setup}
\begin{figure}[bt]
\includegraphics[width=\columnwidth]{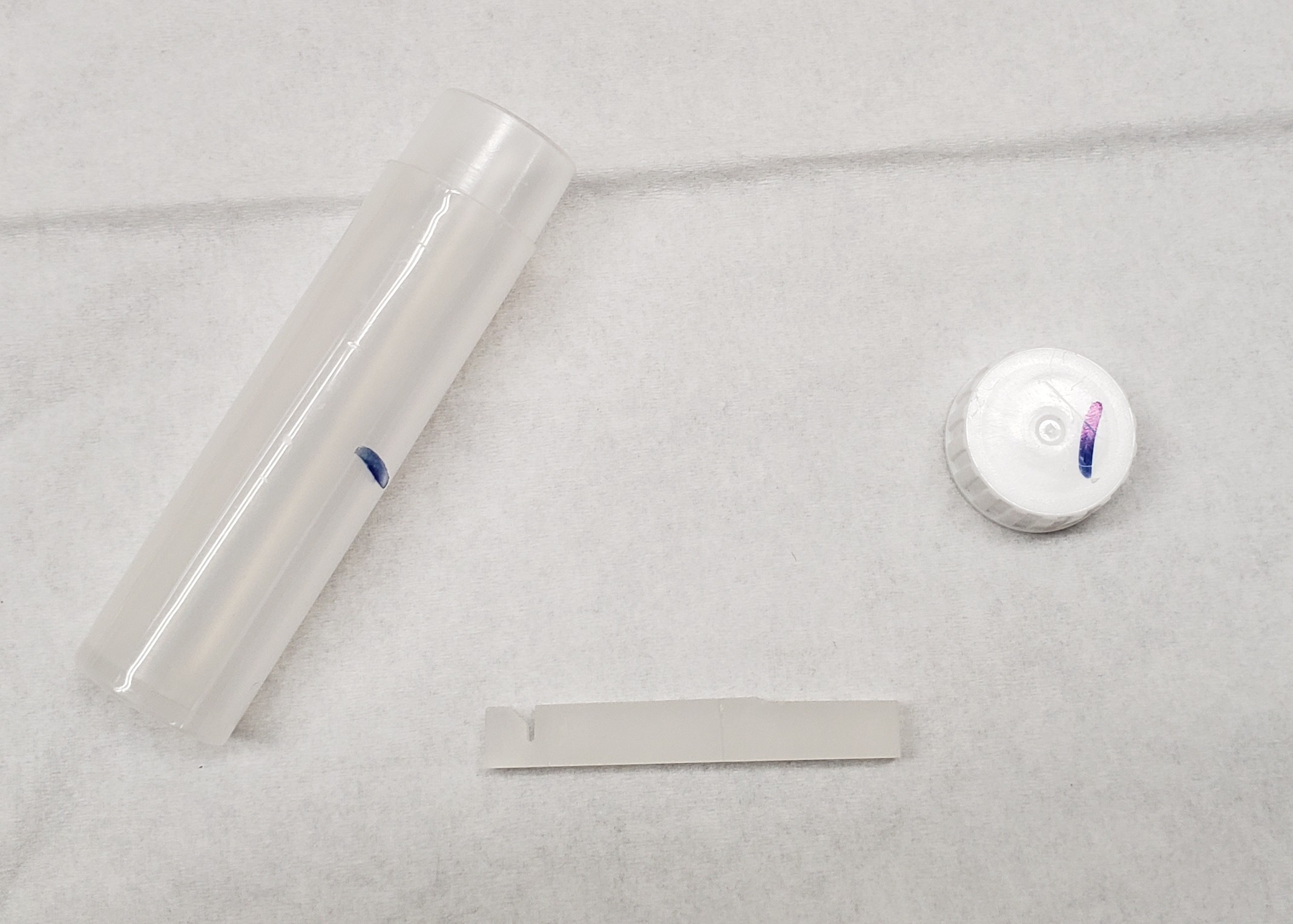}
\caption{\label{fig:sapphire_sample}
The sapphire sample is shown (middle) together with its PE counting vial (left) and snap lid (right).
}
\end{figure}

A $2.5482(3)\gram$ sample of sapphire, received in 2015 from Saint-Gobain Crystals, New Hampshire 
(now Luxium Solutions), was obtained from a larger crystal that was itself saw cut, ground, and annealed at Saint-Gobain. 
The sample designation is G3.
The sample was subsequently cut to the largest size that would fit inside of an activation vial by the University of Massachusetts Amherst glass shop.
The sapphire sample has approximately cuboid shape with side dimensions of $\rm (3.8-4.0)\; mm\times (3.6-4.1)\; mm\times 34.5\; mm$, as shown in Fig.~\ref{fig:sapphire_sample}.
Prior to shipping to the University of Alabama (UA), it was washed in an ultrasonic bath of acetone and then ethanol, for ten minutes each. 

NAA determines the summed contributions of surface and bulk impurities. The measurement reported here states contaminant concentrations normalized by mass. 
To suppress contributions from surface impurities, the sample was carefully cleaned at Pacific Northwest National Laboratory (PNNL) before activation at MIT. After that the sample was only handled in clean room environments or a fume hood.

The precision cleaning procedure developed at PNNL employed a series of sequential ultrasonic baths in ultrapure solutions of 2\% micro-90 detergent, 50\% $\rm HNO_3$, 50\% HCl, and water.
The need for careful and thorough chemical cleaning of the surface was evident from a 2016 measurement of Saint-Gobain sapphire sample A1 (same material as G3 but different surface treatment). 
This assay, also by NAA, did not use an effective cleaning protocol and, as a result, measured \utwothirtyeight\ concentrations in the parts-per-billion range. PNNL measured the leachates from the cleanings using inductively coupled plasma mass spectrometry. 
This ensured that iterative sonication leachate levels had reached a baseline before submitting the cleaned sample for NAA. 
A 2017 NAA re-assay of the PNNL-cleaned sapphire sample G3 yielded only upper limits for $\rm ^{232}Th$ and $\rm ^{238}U$ in the ppt-range. 

After cleaning, samples were packaged in similarly cleaned polyethylene (PE) bottles, which are required to separate the sapphire from the environment inside the research reactor. 
Pre-activation sample handling was performed in a class 500 clean room at UA. 

After sample return to the respective counting labs, the sample was removed from its co-activated PE bottle and transferred into a clean cylindrical PE counting bottle with $\rm 11.7\; mm$ diameter and $\rm 54.5\; mm$ height. 
Post activation, sample cleanliness is no longer an issue, as activities due to short-lived nuclides are not naturally occurring. Typical sample-induced counting rates are of order kHz, much higher than detector background.
The post activation handling, instead, focuses on safety and containment of the activity.

The sapphire sample was activated at the MIT research reactor on May 22, 2024 for a duration of $8.28000(28)\hour$. This irradiation was immediately followed by the activation of two standards, consisting of NIST reference material SRM 1633b (\flyash) with masses of $0.4059(3)\gram$ and $0.5083(3)\gram$. The standards were activated for only 187(1)\seconds\ to avoid creating dangerously high activities.
The 2PH1 sample insertion facility was used for all samples as it offers the highest thermal neutron flux. 
The sapphire and one \flyash\ sample were sent to TUNL for \gammagammacountingnohyphen. To determine the neutron flux, and to ensure analysis consistency between the two facilities, the other \flyash\ sample was sent to UA for single \gammacounting. In all cases, the end of activation established time zero for the following analyses. The fly ash samples were delivered to both analysis labs on May 24, 2024.
After concluding counting of the sapphire sample at TUNL, it was sent to UA to assay long-lived activities and establish analysis consistency. The sample was received at UA on August 20, 2025, 90 days after the end of activation.

The TUNL \gammagammacounting\ setup consists of two ORTEC $p$-type coaxial high-purity germanium (HPGe) detectors. 
The detectors have nominally identical geometries: 88\mm\ crystal diameter, 49\mm\ crystal length, and a 2.54 mm thick nickel-plated magnesium window. The two detectors are herein referred to as Ge-A and Ge-B and have a resolution at 122 keV of 1.8 and 1.1 keV full width at half maximum (FWHM), respectively. The Ge-A and Ge-B resolution at 1332 keV is 2.7 and 2.0 keV FWHM, respectively. 
The detectors are surrounded by passive shielding consisting of 2\cm\ of Cu and 15\cm\ of Pb. A 3D-printed sample holder is utilized to keep the two HPGe detectors collinear and to center the sample between the two detectors. 
This setup was previously used for studies of $\upbeta\upbeta$ decay to excited final states \cite{Kidd2014, Finch2015} at the Kimballton Underground Research Facility. It was relocated to ground level at Duke University prior to the present work.

The data acquisition system uses the Mesytec MDPP-16 digitizer and associated MVME acquisition software. 
The trapezoidal filter of the digitizer is optimized for energy resolution and low trigger threshold, at the expense of timing resolution.  
A 780\ns\ coincidence time window was employed to avoid accidental coincidences with minimal loss of coincident events. This choice was validated with coincident $\gamma$ rays from calibration sources described below.

The samples were counted with the goal of keeping the trigger rate of each detector below 10\kHz\ ($\lesssim 20\%$ dead time). This was accomplished by counting both the \flyash\ and sapphire samples 5\cm\ from each detector face (10\cm\ Ge-Ge separation distance) before counting the samples at a 1\cm\ distance (2\cm\ Ge-Ge separation distance), after the samples had sufficiently cooled. At the 5\cm\ displacement, the \flyash\ sample was counted continuously for 48\hour\ starting on May 25, 2024; the sapphire sample was counted continuously for 24\hour\ starting on May 27, 2024. The detector distance was then reduced to 1\cm\ and the \flyash\ was then counted for 24\hour, followed by 3\days\ of sapphire sample counting. Following this initial counting, the samples were counted continuously for one month, alternating 4\days\ of sapphire counting and 3\days\ of \flyash\ counting. This procedure allowed for most NAA products to be monitored across several half-lives.  

The HPGe detector efficiencies were determined using commercial $\gamma$-ray calibration sources from Eckert \& Ziegler. \cosixty\ radioactivity was used in form of a button source.
A so-called ``scatterless'' source contains \mnfiftyfour, \cofiftyseven, \znsixtyfive, \yeightyeight, \cdoneonine, \snonethirteen, \csonethirtyfour, \csonethirtyseven, \ceonethirtynine, and \amtwofortyone, sandwiched between two thin aluminized Mylar sheets. 
These sources were chosen as they span a wide range of energies, from 60--1836\keV, and emit coincident $\gamma$
rays (\cosixty, \yeightyeight, and \csonethirtyfour). Source calibrations were performed before and after sample counting, in both the 1\cm\ and the 5\cm\ counting geometry. 
Data taken with these activity-calibrated sources served to tune and verify the validity of the detector response model, described in Sec.~\ref{sec:mc}. 

The system dead time was determined using the  procedure recommended by the manufacturer of the digitizer. A histogram was filled with the time difference between all successive events. 
For a Poisson process, this histogram should exhibit an exponential distribution. The digitizer firmware requires 9\micros\ to process each pulse, resulting in a loss of events below this threshold. The histogram is fit with an exponential function above 10\micros; this function, projected down to 0\micros, provides an estimate the number of events lost below the 9\micros\ cutoff. Additionally, the digitizer firmware provides a pileup flag if multiple pulses are detected within the trapezoidal-filter window. 
These flagged events are vetoed from the analysis and counted as an additional source of inefficiency. The \gammagamma\ coincidence dead time was calculated using the above procedure, but requiring that both detectors were not dead. That is, for uncorrelated triggers, the coincidence live time fraction is the product of the two individual detector live time fractions.  

Two $^{60}$Co sources of differing strengths were used to measure the single and coincidence efficiency at HPGe trigger rates of 1\kHz\ and 10\kHz. The two sources produced detector efficiencies consistent to better than $2\%$ at both trigger rates, indicating the live-time correction is accurate to better than $2\%$, up to a 10\kHz\ trigger rate. 
As such, we assume a $2\%$ systematic uncertainty on the detector live time, which is sub-dominant to the uncertainty associated with the detector response model presented in Sec.~\ref{sec:mc}. 
We note that the activity of each $^{60}$Co source was only certified to $1.2\%$, so the $2\%$ systematic is likely an overestimate of the true uncertainty associated with the live time estimation.

\section{\label{sec:mc}Detector response modeling}
\begin{figure}
\subfloat[Sapphire]{\includegraphics[width=\columnwidth]{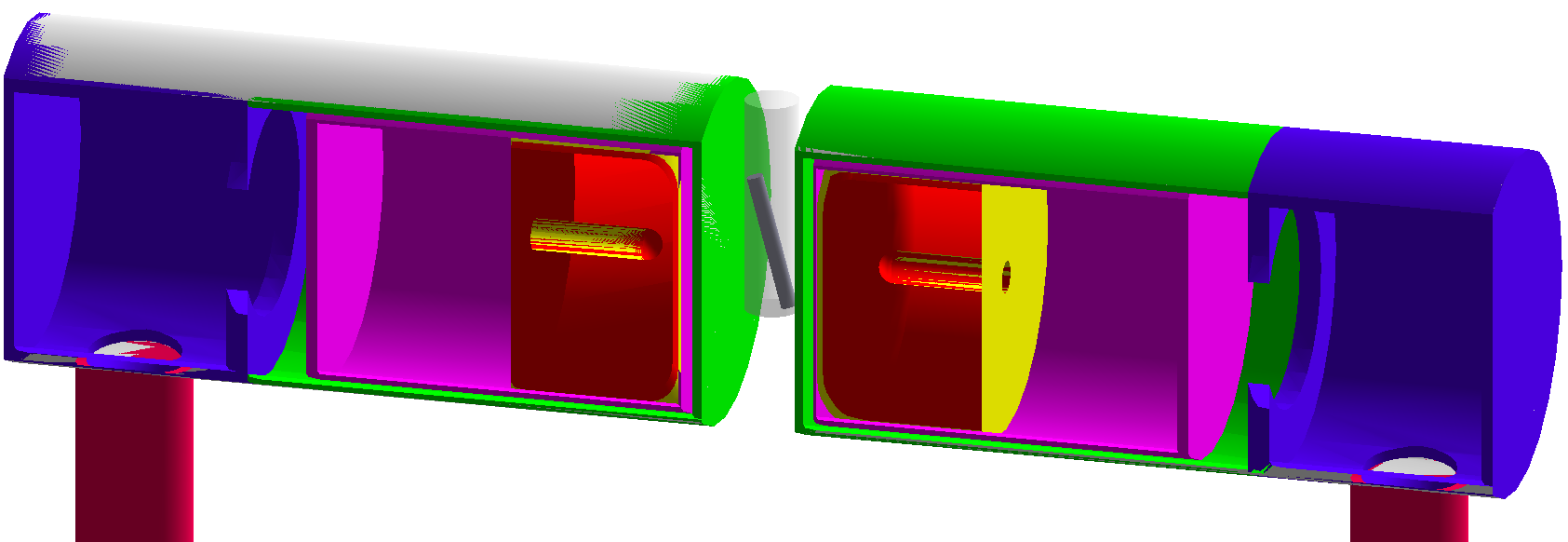}}\\
\hspace{\fill}
\subfloat[Scatterless source.]{\includegraphics[width=0.45\columnwidth]{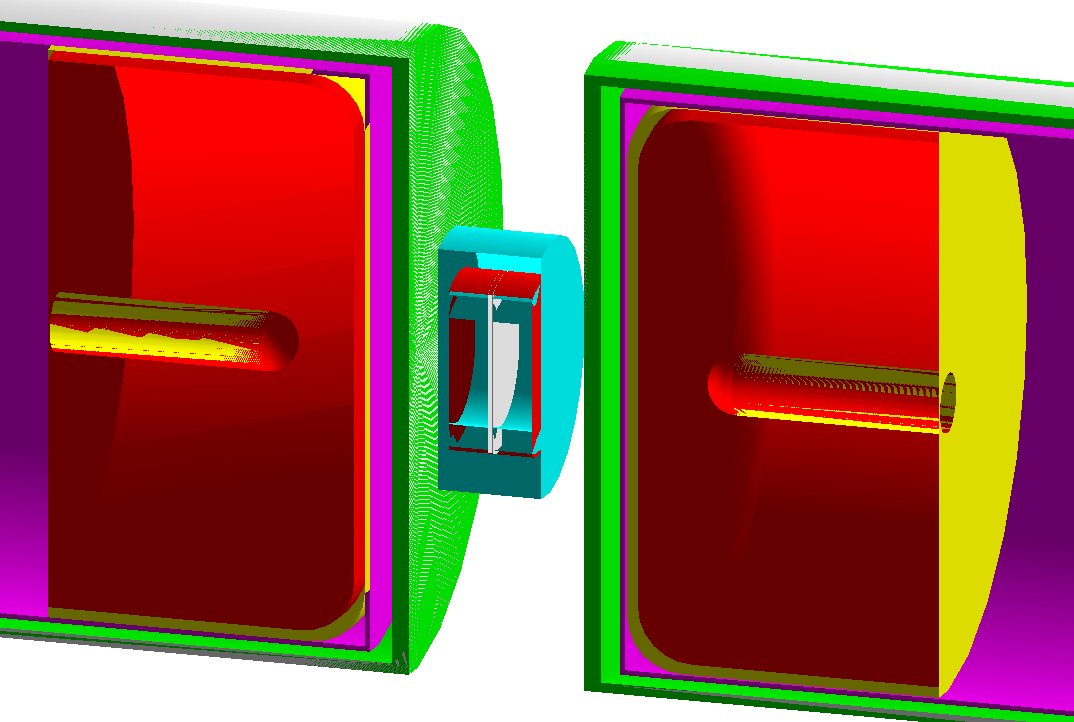}}
\hspace{\fill}
\subfloat[Button source.]{\includegraphics[width=0.45\columnwidth]{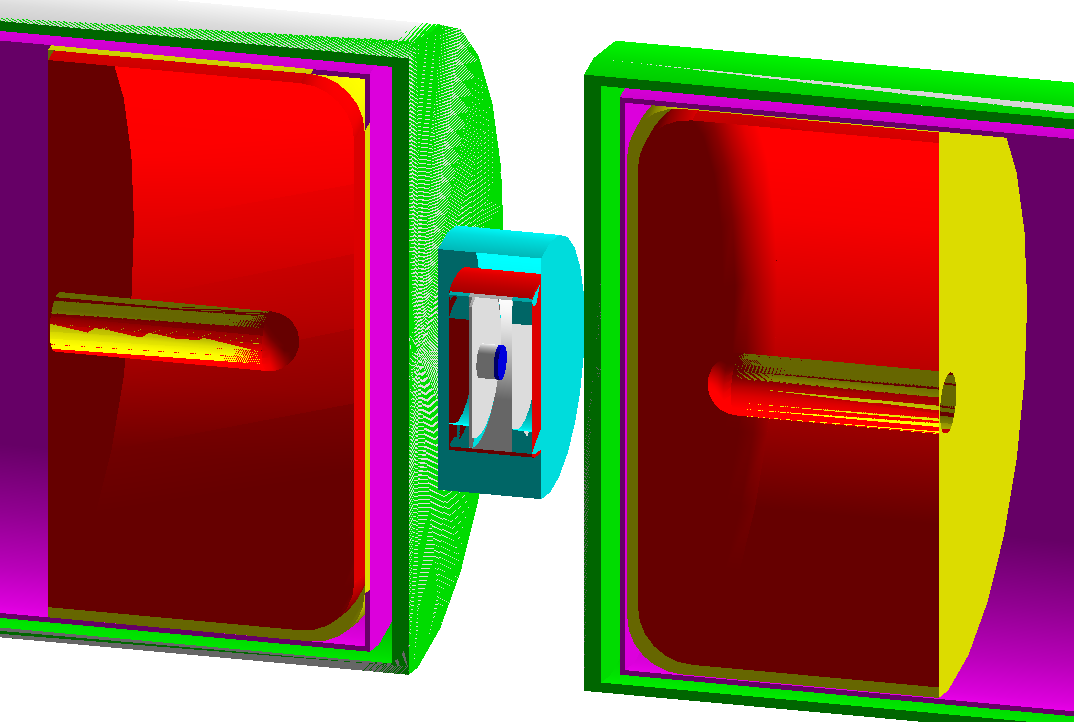}}
\hspace{\fill}\\
\caption{\label{fig:simgeo}Simulated detector geometry. Detectors separated by 2\cm. The sapphire is shown inside a sample counting bottle.
Because of the size mismatch between sapphire sample and counting bottle, the location and orientation of the sample were not well defined. Simulations were performed for various orientations. The top panel shows the sample leaning diagonally inside the bottle; its most likely orientation.
The two calibration source geometries (gray) are shown inside a source holder with different thickness spacers. The sample holder, which sets the separation between the detectors, is not modeled.}
\end{figure}

A custom Geant4 \cite{GEANT4:2002zbu} 10.7.4-based simulation was developed to model the response of both HPGe detectors to activated fly ash and sapphire samples, and the calibration sources. 
It relies heavily on well-tested simulations used to model the response of the Ge II and III detectors at UA~\cite{ua_ge_sim_2019}, and the Ge IV detector in the Black Hill Underground Campus of SURF. It was developed from the simulation code used to estimate the sensitivity of this type of coincident Ge detector measurement in \cite{tsang_2021}. 
The geometry of the Ge crystal, contacts, housing, insulation, and detector cryostat were provided by the manufacturer ORTEC for each detector. The thickness of the outside and core dead layer thickness are nominally (as of August 11, 1997) 700\microm\ and 0.3\microm , respectively. 
In addition, the relevant geometry of the sample and calibration source holders was incorporated into the simulation. The geometry for estimating the efficiency of the detectors to the sapphire sample, with the detectors spaced by 2\cm, is shown in Fig.~\ref{fig:simgeo}. The 3D-printed sample holder, which sets the spacing between the two detectors, is not included in the model as it does not impinge on the line-of-sight between any radioactivity and the Ge crystal of either detector.

The physical characteristics of the HPGe detectors---in particular, their dead layer, the geometry of their germanium crystals, and the separation between the detectors---were known with limited precision. 
The scatterless source with mixed isotopes, with a broad spectrum of activity, was used in the simulation to adjust the parameters related to these unknowns. 
A dead layer for both detectors of about 1.6\mm, a true separation of 2.6\cm, and a source offset by 1\mm, were found to make the source-derived detector efficiency more compatible with that derived from the simulation. 
\begin{figure}
  \includegraphics[width=\columnwidth]{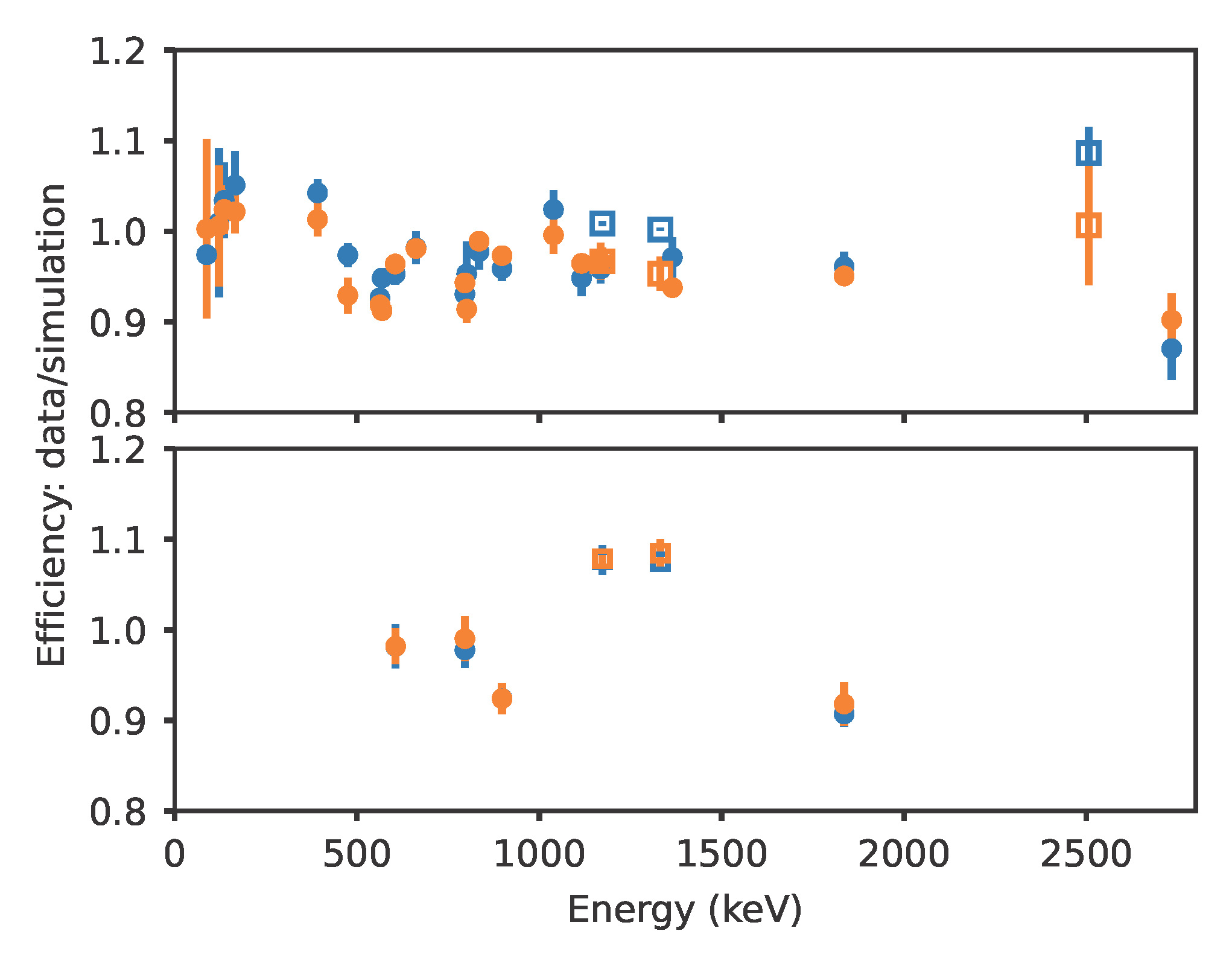}
  \caption{\label{fig:mcefficiency}Ratio of calibration source-derived detection efficiency to simulation-derived detection efficiency for prominent $\gamma$ rays from the scatterless source (see text for a list of nuclides) and from the \cosixty\ source (open squares). Ge-A in orange; Ge-B in blue. Data points are averages over several calibration runs with their RMS errors. Top are for single $\gamma$ rays. Bottom includes coincident pairs of the most prominent $\gamma$ rays from \cosixty, \yeightyeight, and \csonethirtyfour. Both permutations of the coincidence condition are shown in the lower figure: like pairs of orange points and like pairs of blue points are not independent measurements; the analysis does not depend on the choice of tagging detector.}%
\end{figure}

The residual discrepancy between the efficiency derived from calibration source data and that derived from the simulation is shown in Fig.~\ref{fig:mcefficiency}. Only scatterless source data that was not used to tune the simulation (independent detector calibration data) is shown.  Several \cosixty\ source runs, with more complex source geometry when compared to the scatterless source, are also shown. The agreement between the source data and simulation validates a 10\% systematic uncertainty we apply to single \gammaray\ efficiencies derived from this simulation.

The coincident detection efficiency derived from the source data is compared to that derived from the simulation in Fig.~\ref{fig:mcefficiency}, bottom. Pairs of the two most prominent \gammaray\ transitions---of \cosixty, \yeightyeight, and \csonethirtyfour, separately ---were analyzed, requiring coincidence with each other (blue and orange in the figure).
These analysis differ slightly, depending on which detector is designated as the tagging detector; the same signal events are analyzed, only the background events may differ. The lack of variation between the permutations in Fig.~\ref{fig:mcefficiency}, bottom, indicates that this analysis is insensitive to the choice of which detector serves as the tagging detector. In general, agreement is observed that validates a 10\% systematic uncertainty on coincident \gammaray\ efficiency derived from the simulation.

In addition to general errors arising from evaluating sample efficiency with this simulation, there are several specifically associated with simulating the sapphire sample. 
The dimensions and density of the sapphire sample, and its position and orientation within the counting vial, were all varied in the simulation. 
The dimensions and density of the sample are well known, resulting in small contributions to the overall uncertainty. The uncertainty in the position of the sample dominated the total error on the single \gammaray\ counting efficiency. In contrast, the orientation of the sample dominated the error on the coincident \gammaray\ efficiency.  
The position of the sample within the counting vial altered the simulated sapphire single \gammaray\ detection efficiency within a $\pm7\%$ (relative) range for either detector. 
The orientation of the sample, tilting towards one detector or the other, increased the simulated sapphire coincident \gammaray\ detection efficiency by up to 3\% (relative). This uncertainty is subdominant to, and to some extent correlated with, the detector response model uncertainty as measured with the calibration sources.

\section{\label{sec:analysis}Data analysis and results}
Material assay by means of NAA is based on the measurement of radioactivity induced by the radiative capture of neutrons on target nuclides of interest. In case of uranium, the reaction $^{238}$U$(\neutron,\upgamma)^{239}$U is used. The capture product decays into $\rm ^{239}Np$ with a half-life of 23.45\minute, which then \betadecay s into $\rm ^{239}Pu$ with a half-life of 2.356\days. 
This is long enough to allow shipment from a research reactor to the counting laboratory.
Knowing the sample mass, the time and duration of the irradiation, and the neutron flux and capture cross sections, allows the conversion of measured activities into the concentrations of the target elements. The details of this conversion are described in Refs.~\cite{tsang_2021, leonard_2017}.

\subsection{Neutron flux}
\newcommand{\Phith}{\ensuremath{\Phi_\text{th}}}
\newcommand{\Phiepi}{\ensuremath{\Phi_\text{et}}}
\newcommand{\Phifast}{\ensuremath{\Phi_\text{f}}}
The UA fly ash counting data was used to determine that the thermal \Phith, epithermal \Phiepi, and fast neutron flux \Phifast, for this MIT reactor activation, were%
\NewDocumentCommand\mkwidthof{O{\raggedright}mm}{%
 \parbox[b]{\widthof{#2}}{#1#3}}%
\NewDocumentCommand\mmkwidthof{O{\raggedright}mm}{%
 \mkwidthof[#1]{\ensuremath{\displaystyle#2}}{\ensuremath{\displaystyle#3}}}%
\begin{eqnarray*}
    \Phith   & = & \mathrm{(4.905\pm 0.088)\cdot 10^{13}\,\frac{\text{neutrons}}{\text{cm}^2\cdot \text{s}}} \\
    \mmkwidthof{\Phith}{\Phiepi}  & = & \mathrm{(1.50\pm 0.12)\cdot 10^{13}\,\frac{neutrons}{cm^2\cdot s}} \\
    \mmkwidthof{\Phith}{\Phifast} & = & \mathrm{(7.39\pm 0.88)\cdot 10^{12}\; \frac{neutrons}{cm^2\cdot s}}
\end{eqnarray*}
A point-wise systematic error of 2.8\% has been added in quadrature to the extrapolated fly ash activities to achieve $\chi^2=0.99$ for three degrees of freedom between the observed and modeled activities. Average neutron capture cross sections were taken from the JENDL-4.0 library~\cite{jendl-4.0_2012}. A detailed description of an equivalent neutron flux determination can be found in Ref.~\cite{leonard_2017}.

\subsection{Data analysis} \label{data_analysis}
Single \gammaray\ spectra, not requiring a time coincidence, were analyzed separately for detectors Ge-A and Ge-B. 
Both detectors provide independent event triggers. The single-event distributions contain no linkage between the two detectors and represent different photon populations. They are, therefore, independent of each other. These independent data sets are combined by means of a weighted average, to reduce uncertainty. 
The simulations of the two detectors were tuned separately using radioactive source data. 
We, therefore, assume that their simulation uncertainties are independent of each other. The systematic uncertainty was included in the weights during averaging.

For the \gammagamma\ coincidence analysis, events in one detector were selected with a coincidence condition applied to the other; requiring the presence of an event within a predefined energy region and coincidence time window in the second, the tagging detector. 
The energy regions were determined from the tabulated $\gamma$ cascade energies  of the activation products. They were defined as the total absorption peak $\pm 3 \sigma$ (energy resolution).
The full absorption peak of photon 1 was fit in data collected with Ge-A, with Ge-B acting as the tagger, required to fully absorb coincident photon 2. Requiring photon 2 to be absorbed by Ge-A and tagging photon 1 in Ge-B selects different photons. These two data sets are statistically independent and were combined by averaging. However, the same photons are selected when merely exchanging the roles of the fit with that of the tagging detector. Rates resulting from this permutation of roles were not included in the average.

The total absorption peaks were modeled as Gaussian, superimposed on a linear background. For short-lived, high-activity products, a low-energy tail and a step in the background at the centroid were added to the model, as detailed in Ref.~\cite{arnquist_2023}. For
\gammagamma\ coincidences of relatively long-lived, low-activity products, only a Gaussian was required to model the data. 
For the sapphire sample, so few events satisfied the coincidence condition for \nptwothirtynine\ that an estimate of the observed \nptwothirtynine\ activity was derived from integrating the residual spectrum over a $\pm 3 \sigma$ region of interest. The background contribution in this region was estimated from the integral over $6\sigma$.
This approach was validated against the fit-based analysis by using the \nptwothirtynine\ \gammagamma\ coincidences observed in the fly ash data.

The time dependence of the counting rates was modeled with an exponential function, fixing the half-life to that of the nuclide identified through the choice of full absorption peak. The analysis has high selectivity because it is double differential in energy and time.
Extracted counting rates, extrapolated to the end of activation, along with the detector response model-derived efficiency and measured neutron flux, were then used to quantify elemental concentrations in the fly ash and sapphire samples. 

\newcommand{\fsample}{\ensuremath{c_\text{s}}}
\newcommand{\msample}{\ensuremath{m_\text{s}}}
\newcommand{\mfa}{\ensuremath{m_\text{a}}}
\newcommand{\ffa}{\ensuremath{c_\text{a}}}
\newcommand{\Rs}{\ensuremath{r_\text{s}(0)}}
\newcommand{\Rfa}{\ensuremath{r_\text{a}(0)}}
\newcommand{\epsilonfa}{\ensuremath{\varepsilon_\text{a}}}
\newcommand{\epsilons}{\ensuremath{\varepsilon_\text{s}}}
\newcommand{\tfa}{\ensuremath{d_\text{a}}}
\newcommand{\tsample}{\ensuremath{d_\text{s}}}
Because this study uses a novel approach and a detection system differing from previous studies, we decided to also use an alternative approach to the conversion of counting rates into element concentrations. This approach is less reliant on the detector response modeling. Both the standard fly ash material and the sapphire were exposed to the same neutron environment and counted with the same detector setup, with similar geometry. 
The ratio of counting rates eliminates the dependence on measured neutron flux and capture cross sections. The 
elemental concentration in the sample, \fsample, was calculated from the ratio of sample to fly ash count rates, extrapolated to the end-of-activation $\Rs\big/\Rfa$, as%
\begin{eqnarray*}%
    \fsample & = & \frac{\mfa \cdot \ffa}{\msample} \cdot \frac{(1-e^{-\tfa/\tau})}{(1-e^{-\tsample/\tau})} \cdot \frac{\Rs}{\Rfa} \cdot \frac{\epsilonfa}{\epsilons},%
\end{eqnarray*}%
where \ffa\ is the elemental concentration in the fly ash sample, \msample\ and \mfa, and \tsample\ and \tfa\ are the masses and duration of the activation of the sample and fly ash, and $\tau$ is the mean lifetime of the nuclide of interest.
The ratio of the relative efficiency of the fly ash to sapphire samples, $\epsilonfa\big/\epsilons$, is a correction of order one that accounts for the slight difference in sample geometry, as determined by the detector response model.

This so-called ``fly ash'' 
method is applicable only to Cr, Th, and U, whose activation products were observed in both the fly ash and sapphire samples. 
The analysis of all other elements relies entirely on the measured neutron fluxes and the neutron capture cross sections, as in \cite{leonard_2017}.

\subsection{Quality checks \label{sec:quality_checks}}
\begin{figure}[bt]
\includegraphics[width=\columnwidth]{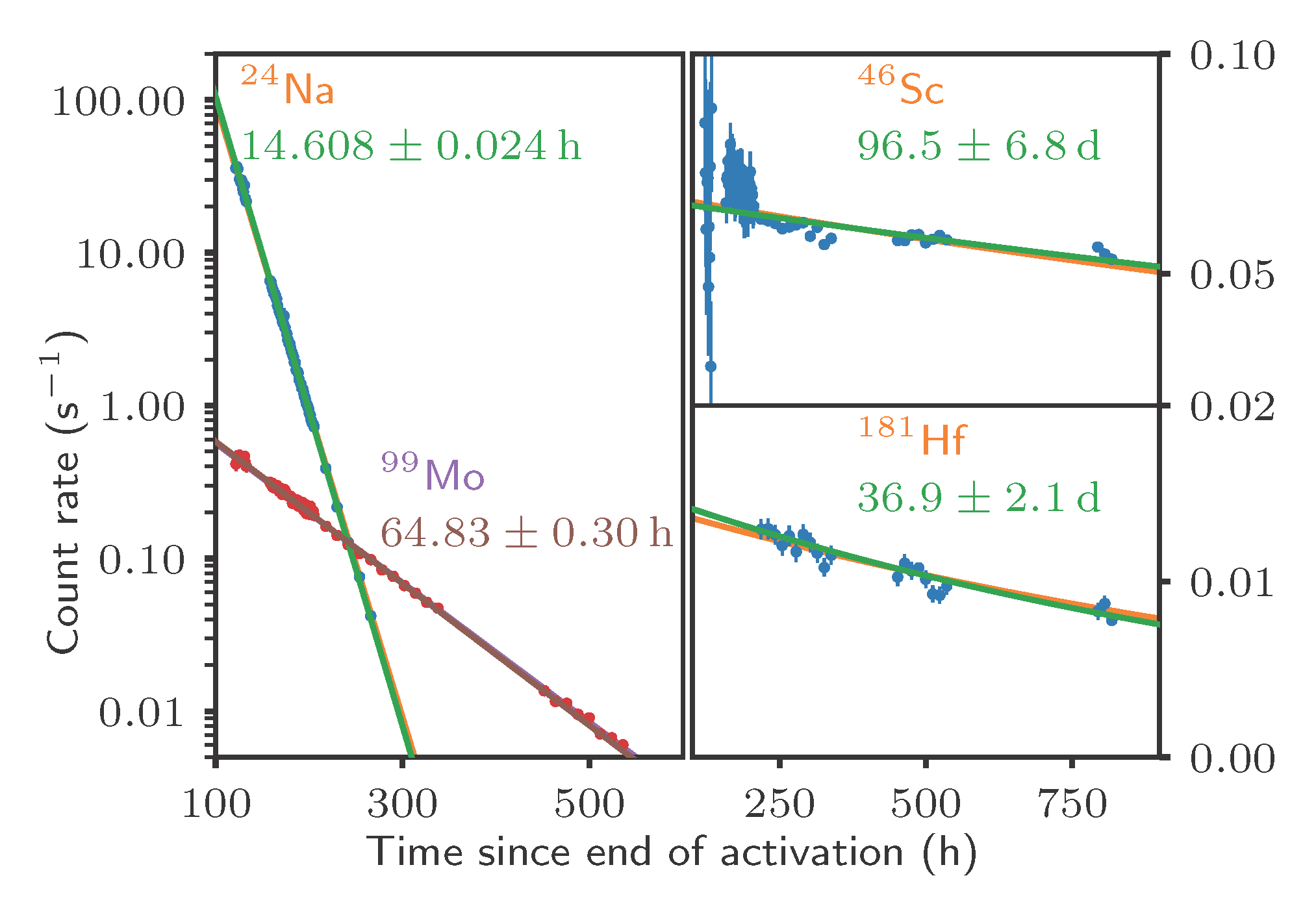}
\caption{\label{fig:sapphire_time_fits} 
Time dependence of the observed count rates for four coincident photopeaks associated with \natwentyfour, \scfortysix, \moninetynine, and \hfoneeightyone\ present in the sapphire sample. 
An exponential decay model, both with fixed (indicated by the isotope) and free half-life parameter (shown with uncertainty), are shown after a least squares fit to the normalization of the data points. The lines conform to the color coding of the text.
}
\end{figure}
Because this was the first use of the described method, care was taken to validate results. The following tests were performed to ensure the consistency of the method with previous measurements:
\begin{enumerate}
    \item To test the compatibility of the detector response models employed at UA and TUNL, we determined the reactor neutron flux by counting the fly ash sample at UA and then used this flux calibration to derive the common elemental concentrations from counting the TUNL fly ash sample.\label{test-1} \\
    The TUNL-data derived concentrations of Na, Cr, Fe, Se, As, Ba, Th and U were compared to their NIST certified values. Only single \gammacounting\ data was used for Cr and Th due to the lack of a suitable \gammaray\ cascade. 
    For Se, only \gammagammacounting\ results were obtained due to high background in the individual detectors.
    The ratio of TUNL results to certified values, averaged over all elements, was found to be $1.084\pm0.031$ for single \gammacounting\ and $1.034\pm0.032$ for \gammagammacountingnohyphen. 
    The Ge-A to Ge-B derived ratios of concentrations in sapphire have a weighted average of $1.034\pm 0.016$ when averaged over Sc, Cr, Co, Mo and Hf, demonstrating consistency between the detector response models.
    \item In 2017, elemental concentrations of Sc, Cr, Co, Mo and Hf were determined at UA  by activating and single \gammacounting\ the same Saint-Gobain G3 sapphire sample.  In 2017 the 1PH1 sample insertion facility was used. 
    It has a different neutron energy spectrum compared to 2PH1\, used for this analysis. 
    TUNL single-$\upgamma$ and \gammagammacountingnohyphen\ results were compared with the 2017 UA data.\label{test-2}\\
    The average ratio of the concentrations of these elements derived from \gammacounting\ was found to be $0.95\pm0.06$. Again excluding Cr, the average \gammagammacountingnohyphen-derived TUNL to UA concentration ratio was $1.00\pm 0.07$.\\
    The corresponding averages for TUNL to UA 2024 concentrations were found to be $1.13\pm 0.07$ and $1.18\pm 0.10$, respectively. Mo was not included in the 2024 comparison as it had decayed by the time counting had commenced at UA.
    \item For Cr in Saint-Gobain G3 sapphire, the elemental concentration evaluated via the measured neutron flux, capture cross sections and detector response model-derived efficiency was compared with that resulting from the fly ash method and single \gammacounting\ at TUNL.\label{test-3}\\
    Averaging over both Ge detectors, we found a concentration ratio for the two methods (model over fly ash method) of $1.15\pm 0.08$.
    \item Fig.~\ref{fig:sapphire_time_fits} demonstrates the ability to properly identify elemental species. Using sapphire \gammagammacounting\ data we extract the time dependence of the \natwentyfour, \scfortysix, \moninetynine, and \hfoneeightyone\ event rates. \\
    The data is modeled with an exponential decay; both with the half-life fixed to that of the expected isotope (taken from \cite{Na24_half-life,Mo99_half-life,Hf181_half-life,Sc46_half-life}), and with the half-life as a free parameter. The average ratio of the expected-to-free half-lives is $0.9774\pm 0.0015$. The rate uncertainty due to sample positioning is not included in the least square fits of this model to the data and, therefore, not reflected in the uncertainty of the extracted half-lives. 
    Allowing the normalization of the various distinct groups of data points to float, each group representing data taken with the sample in precisely the same position, improves the agreement with the data, and the agreement with the expected half-life. However, this model was not generally applied to all the data since the individual normalization of the separate positions must be averaged to obtain the initial activity of the sample; the variation is accounted for by the application of the more robust position derived uncertainty from section \ref{sec:mc}.
    We consider the observation that measured and tabulated half-lives agreed, on average, within 2.3\% as evidence that the elemental species can be properly identified by this type of decay time analysis.
\end{enumerate}
From these tests, we conclude that the new method gives results that are comparable to previous analyses. This also holds for the conversion of event rate into concentrations by the fly ash method, largely avoiding the dependence on cross sections and detector response model.

\subsection{Analysis results}
\begin{table*}[htb]
    \centering
    \begin{tabular}{l|l|l|l|l|l|l|l|l}
        Nucl. &  Target &  \multicolumn{3}{c|}{Concentration [pg/g]} \\ 
              &    & single-$\gamma$ & $\gamma$-$\gamma$-coin & UA 2017 & $\gamma$/UA 2017 & $\gamma$-$\gamma$/UA 2017 & $\gamma$/UA 2024 & $\gamma$-$\gamma$/UA 2024\\
        \hline\hline
        $\rm^{46}Sc$ & Sc & $345 \pm 25$ & $358 \pm 36$ & $375 \pm 39$ & $0.92 \pm 0.12$ & $0.96 \pm 0.14$ & $1.17\pm 0.15$ & $1.22\pm 0.17$\\
%
        $\rm^{51}Cr$ & Cr & $27500 \pm 2000$ & & $32540 \pm 3340$ & $0.85 \pm 0.11$ & & $1.09\pm 0.14$ &  \\
%
        $\rm^{60}Co$ & Co & $278 \pm 20$ & $278 \pm 29$ & $193 \pm 20$ & $1.44 \pm 0.18$ & $1.44 \pm 0.21$ & $1.17\pm 0.15$ & $1.17\pm 0.17$ \\
%
        $\rm^{99}Mo$ & Mo & $1.093 \pm 0.083^*$ & $1.10 \pm 0.12^*$ & $1.22 \pm 0.13^*$ & $0.90 \pm 0.12$ & $0.90 \pm 0.14$ &  &  \\
%
        $\rm^{181}Hf$ & Hf & $479 \pm 34$ & $502 \pm 51$ & $520 \pm 53$ & $0.92 \pm 0.12$ & $0.97 \pm 0.14$ & $1.11\pm 0.14$ & $1.16\pm 0.17$  \\
        \hline
        \multicolumn{5}{l|}{Average}  & $0.947\pm 0.061$ & $1.002\pm 0.074$ & $1.131\pm 0.070$ & $1.180\pm 0.098$\\
        \hline
        $\rm^{233}Pa$ & Th & $0.53 \pm 0.32$ & & $-0.88 \pm 0.34$ & & &  &  \\
        & & $(< 1.1)$ FA & & $(< 0.10)$ & &  &  &  \\
%
        $\rm^{239}Np$ & U & & $-2.7 \pm 2.2$ & $5.1 \pm 3.8$ & &  &  &  \\
        & & & $(<1.4) $ FA & $(<11)$ & &  &  &  \\
        \hline

    \end{tabular}
    \caption{Elemental concentrations determined for Saint-Gobain G3 sapphire. Values for Mo, marked with a $^*$, are multiplied by a factor $10^{-6}$.
    Data reported in the single-$\gamma$ column are the weighted average of the Ge-A and Ge-B results.
    The reported uncertainties correspond to that of the weighted average, discussed in section~\ref{data_analysis}.
    Concentrations marked with ``FA'' were obtained using the ``fly ash'' method; all others rely on the detector response-model derived efficiency.  
    Concentrations in the $\gamma$-$\gamma$ coincidence column were obtained using the two independent permutations of the coincidence condition and combining their weighted average (see section~\ref{data_analysis}). 
    In addition, the results of the two independent $\rm ^{239}Np$ cascades were combined by means of a weighted average.
    The Th and U concentration limits, in parenthesis, are based on the Feldman-Cousins method~\cite{F-C_1998}. Errors are reported as the quadratic sum of the statistical and the 10\% systematic error associated with the detector response model. For $\rm^{233}Pa$ and $\rm^{239}Np$, the errors are dominantly statistical. Concentrations reported as ``UA 2017'' were obtained using a different neutron flux and energy distribution than ``UA 2024.'' Various ratios are stated with their uncertainties to demonstrate analysis consistency between TUNL and UA, different reactor neutron energy distributions, and singles and coincidence counting. The line denoted ``Average'' summarizes this comparison. The horizontal divider delineates the difference between analysis consistency checks and reporting of method-dependent sensitivity progress for \thtwothirtytwo\ and \utwothirtyeight. 
    }
    \label{tab:sapphire_results}
\end{table*}
This results section focuses on the core subject, the analysis of $\rm ^{238}U$ by means of \gammagammacountingnohyphen, in addition to measurement of $\rm ^{232}Th$ by single \gammacounting.

\begin{figure}[bt]
\includegraphics[width=8.5cm]{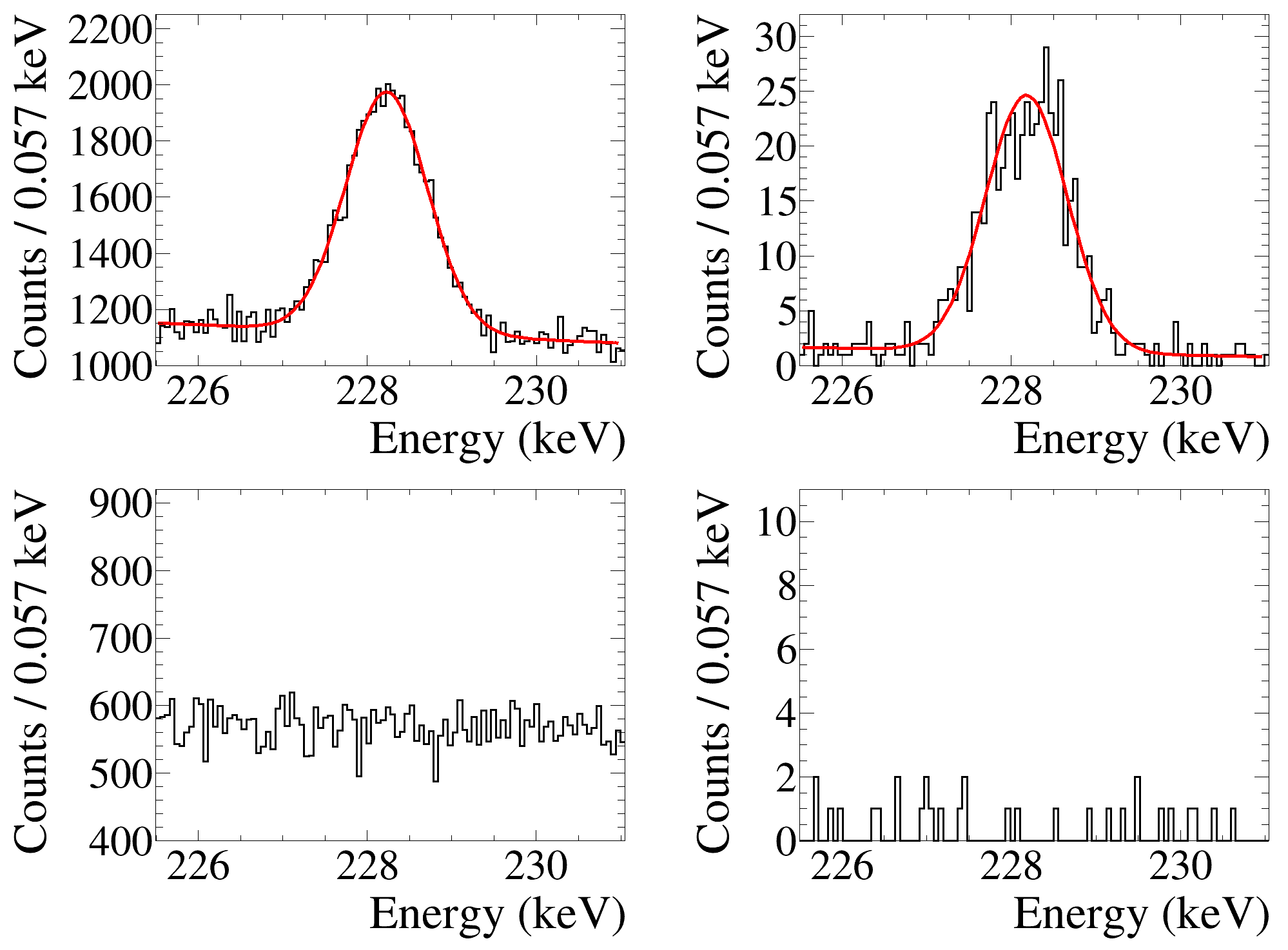}
\caption{\label{fig:energy_spectra} 
Left: single \gammaray\ spectra around 228.2\keV\ measured with Ge-B from fly ash (top) and sapphire sample (bottom). A photopeak from the decay of $\rm^{239}Np$ is expected at this energy~\ref{fig:239np_partial_level_scheme}. Right: \gammagamma\ coincidence spectra measured with Ge-B with the requirement of an energy deposit in Ge-A within $3\sigma$ of the 106.1\keV\ transition. Sensitivity is demonstrated using the fly ash sample. No statistically significant evidence of $\rm^{239}Np$ is observed in the sapphire sample.}
\end{figure}

Fig.~\ref{fig:energy_spectra} demonstrates that \gammagammacountingnohyphen\ has efficiency for the detection of the $\rm ^{238}U$-product $\rm ^{239}Np$. 
The upper row shows the energy distribution of single (left) and coincidence data (right) obtained with NIST fly ash, which contains a known concentration of $\rm ^{238}U$. Clearly, the 228.2\keV\ photopeak is visible in both panels of the upper row. 
The counting efficiency decreases, as expected from the detector response model, when requiring the simultaneous detection of 106.1\keV\ photons with the other detector; while the signal-to-background ratio significantly increases. 
The lower row of Fig.~\ref{fig:energy_spectra} shows the same energy range for data obtained with the sapphire sample. No peak is visible in either the single $\gamma$ (left) or in the \gammagamma\ coincidence (right) spectrum. However, a dramatic background suppression is observed. 

\begin{figure}[bt]
\includegraphics[width=8.5cm]{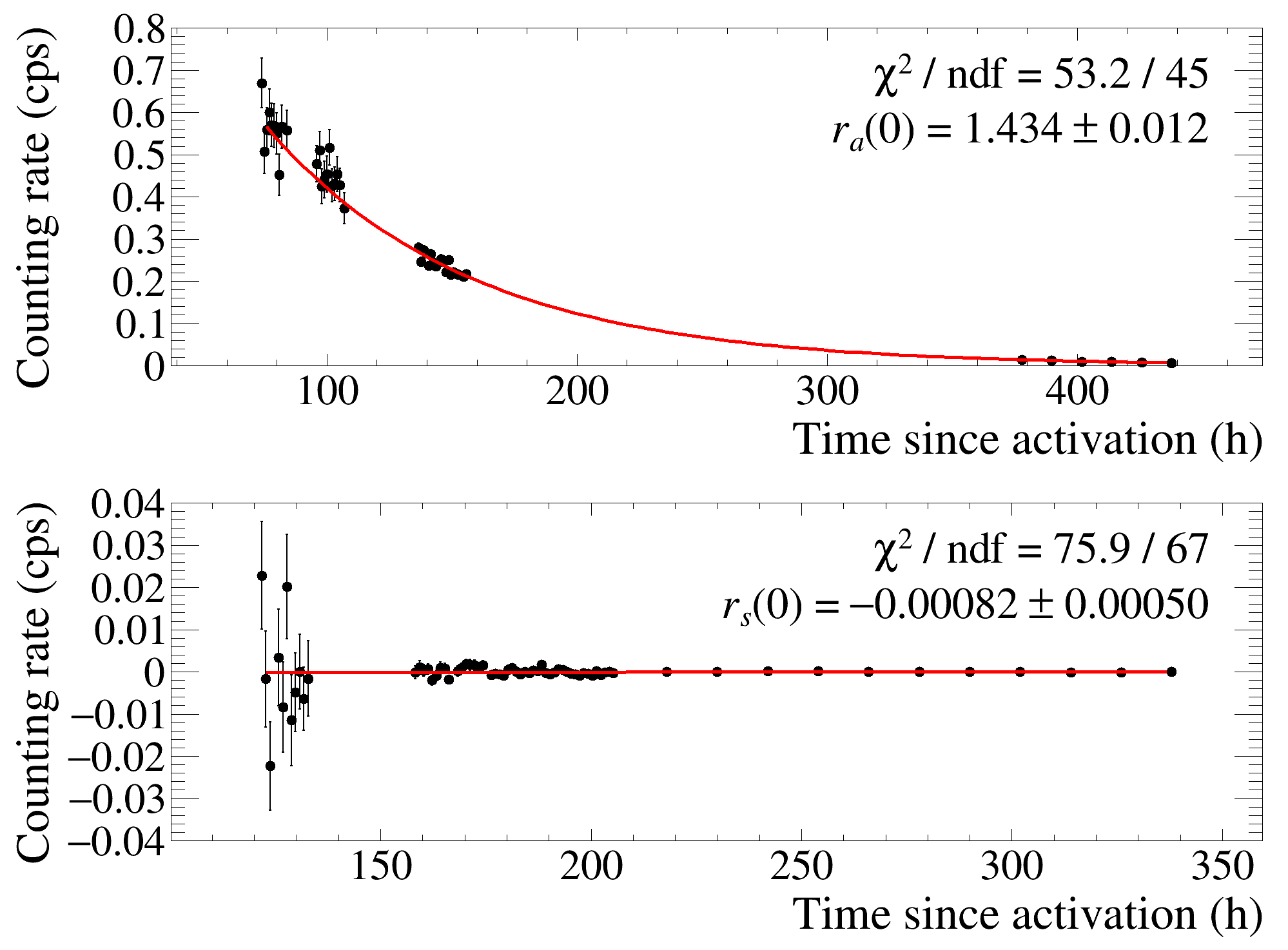}
\caption{\label{fig:np239_time_fits} 
Time dependence of the \nptwothirtynine\ 277.6\keV\ \gammaray\ full absorption peak count rate as observed by Ge-A, requiring coincidence with Ge-B. Top: fly ash, bottom: sapphire. Red line: exponential decay model with the half-life fixed to that of $\rm^{239}Np$. Rates with statistical errors obtained from the fit are shown.
}
\end{figure}
The energy spectra in Fig.~\ref{fig:energy_spectra} represent a snapshot in time, acquired over 1\hour, of the response of Ge-B. The entire counting data are summarized by the normalization of an exponential decay model extracted from a least squares fit to the rate data as a function of time; both shown in Fig.~\ref{fig:np239_time_fits} for the fly ash (top) and the sapphire (bottom). In both panels the half-life has been fixed to that of $\rm ^{239}Np$. The extracted normalization rate from these panels is directly proportional to the activity derived at the end of the activation ($t=0$), which serves as the quantity of interest.

The concentrations of impurities in the sapphire are calculated as weighted means of the results from both detectors, and both cascades in the case of \nptwothirtynine. 
These results are summarized in Tab.~\ref{tab:sapphire_results}. Unless denoted with ``FA'', the concentrations are based on the neutron flux, capture cross section method. Results with ``FA'' are based on the fly ash method.
The primary results for $\rm ^{232}Th$ and $\rm ^{238}U$ are based on the fly ash method which we consider to be the most direct approach, based on comparison to a standard.
Results obtained in a 2017 activation of the same sample are given for comparison. They were derived from single \gammacounting\ only.
The \gammagamma\ coincidence data show no evidence for the presence of $\rm ^{238}U$ in this sample of sapphire. 
The resulting concentration is 1.3 standard deviations negative (unphysical). We use the Feldman-Cousins approach~\cite{F-C_1998} to state a 90\% CL limit in parenthesis. 

The application of the \gammagammacountingnohyphen\ method allowed us to reach ppt sensitivity for $\rm ^{238}U$. We consider this result to be an experimental demonstration of the computational projection published in Ref.~\cite{tsang_2021}.

For $\rm ^{232}Th$ contained in the sapphire, only single \gammacounting\ could be used. However, compared to the 2017 analysis, two detectors were available rather than one. Also, here no evidence of contamination is found. We report the analysis result and Feldman-Cousins limit in parenthesis. 
Fits to the single $\gamma$-spectra of the unobserved 106, 228, and 278 keV peaks in the 2024 TUNL sapphire data resulted in negative rates for all three peaks and both detectors. This behavior is not indicative of a statistical fluctuation. Consequently, we do not derive concentration values from that data.

It is interesting to note that a comparison of 2017 and 2024 results shows physical and unphysical results for Th and U in both combinations. This is the hallmark of low statistics results; they fluctuate around zero.
For enhanced robustness against such fluctuations, and to avoid unphysical results improving the sensitivity (a property of the Feldman-Cousins method), we are basing our final result for $\rm ^{232}Th$ and $\rm ^{238}U$ on the weighted average of the 2017 and 2024 measurements.
We obtain weighted means for Th and U as 
$-0.13\pm 0.23\ppt$ and $-0.8\pm 1.9\ppt$; both only mildly negative.
The resulting Th and U limits are $<0.26\ppt$ and $<2.3\ppt$, respectively. We do not believe that there is a loss in U sensitivity due this averaging; rather that statistical fluctuations have been smoothed via repeated measurements.

\begin{table*}[htb]
  \centering
  \begin{tabular}{l|l|l|l|l|l}
    Supplier & Experiment &  Assay &  Th & U & Ref. \\
             &          &  method  &  [ppt] & [ppt] & \\ \hline\hline
    Single Crystal Technology (Holland) & CRESST &  NAA &  $<33$ & $<40$ & \cite{cresst_1996} \\
    Crystal Systems (USA) & CRESST &  NAA &  $<2.3$ & $<20$ & \cite{cresst_1996}\\ \hline
    RSA (France) & ROSEBUD &  Ge &  $<1200$ & $<400$ & \cite{rosebud_1999} \\
    RSA (France) & ROSEBUD &  Ge &  $<500$ & $<160$ & \cite{rosebud_1999} \\
    Russia & ROSEBUD &  Ge &  $<1200$ & $<400$ & \cite{rosebud_1999} \\ \hline
    Swiss Jewel Company (USA) & EXO-200 & NAA & $30\pm 7$ &  $<25$ & \cite{leonard_2008} \\ \hline
    Marketech Intern. (Taiwan) & Majorana & NAA &  $<21$ & $<300$ & \cite{majorana_2016} \\ \hline
    Hamamatsu (Japan), for R11410-20 & LZ & Ge &  $3500\pm 740$ & $2800\pm 240$ & \cite{lz_2020} \\ \hline
    Hamamatsu (Japan), for R11410-30 & XENON1T & Ge & $1200\pm 310$ &  $3600\pm 200$ & \cite{xenon_2015} \\ \hline
    Precision Sapphire Tech. (Lithuania) & nEXO & NAA &  $410\pm 41$ & $990\pm 99$ & \cite{nexo_assay_2025} \\
    GTAT Corporation (USA), crackle & nEXO & Ge & $<1100$ & $<270$ & \cite{nexo_assay_2025} \\
    GTAT Corporation (USA) & nEXO &  NAA &  $6.0\pm 1.1$ & $<8.9^*$ & \cite{nexo_assay_2025} \\
    Saint Gobain (USA), alumina & nEXO & Ge & $1800\pm 580$ & $<270$  & \cite{nexo_assay_2025} \\
    Saint Gobain (USA), alumina & nEXO & Ge & $6500\pm 1100$ & $2600\pm 520$  & \cite{nexo_assay_2025} \\
    Saint Gobain (USA) & nEXO & NAA & $<0.26$ & $<2.3$ & This work \\ \hline
  \end{tabular}
  \caption{
  Compilation of $\rm ^{232}Th$ and $\rm ^{238}U$ radioassay results found in the literature for sapphire
  and derived from our work. In some cases constraints on $\rm ^{26}Al$, $\rm ^{40}K$, $\rm ^{60}Co$ and $\rm ^{137}Cs$ activities exist but are not shown.
  Crackle and alumina are precursors to the sapphire crystal growth.
  The assay method determines which part of the decay series has been probed. In the case of NAA, $\rm ^{232}Th$ and $\rm ^{238}U$ are measured. In case of Ge counting (Ge), lower chain members are counted and their activities have been converted into concentrations of the chain heads by assuming secular equilibrium. For nEXO data, limits were determined using the Feldman-Cousins approach~\cite{F-C_1998}. 
  The $\rm ^{238}U$ constraint marked with a $^*$ is an exception. The raw result is 2.4 standard deviations negative, resulting in an inflated sensitivity of $\rm <1.8\; ppt$. We decided to not interpret this result as a statistical fluctuation and state the result using the flip-flop method instead, not taking credit for the unphysicality. 
  }
  \label{tab:results_comparison}
\end{table*}

To provide a context for our results, we are comparing them to other sapphire radioactivity measurements found in the literature. Table~\ref{tab:results_comparison} lists all radioactivity measurements we could find. The results attributed to nEXO will appear in a paper that is in preparation. As can be seen in Table~\ref{tab:results_comparison}, there is a broad interest in the radioactivity content of this material. 

\section{Conclusion}
\gammagammacountingnohyphen\ of a neutron activated sample of Saint Gobain sapphire was used to derive 90\% CL purity constraints of $<0.26$\ppt\ and $<2.3$\ppt\ for $\rm ^{232}Th$ and $\rm ^{238}U$, the most stringent Th and U radioassay for sapphire. These results verify the prediction of Ref.~\cite{tsang_2021}.
Clearly, our novel assay method led to a significant sensitivity enhancement for Th and U, compared to what can be found in the published literature. The gain derived from counting \gammagamma\ coincidences might be beneficial for future investigations of trace $\rm ^{238}U$ impurities in sapphire and other materials. We have shown that this method can even cope with substantial low-energy background, often resulting from the presence of chemical impurities that would otherwise limit the NAA sensitivity. The absence of ICP-MS-derived purity measurements, otherwise one of the most sensitive assay techniques, underscores the difficulty in dissolving sapphire in mineral acids.

\begin{acknowledgments}
The authors thank the nEXO collaboration for many stimulating and useful discussions.
This work was supported in part by the U.S.~Department of Energy Office of Nuclear Physics and Office of High Energy Physics under Grant No. DE-FG02-01ER41166, DE-SC0012447, ~DE-FG02-97ER41033, DE-SC0023053, DE-SC0024666,
the Laboratory Directed Research and Development and Early Career Research program at Pacific Northwest National Laboratory (PNNL), a multi-program national laboratory operated for the U.S. Department of Energy
(DOE) by Battelle Memorial Institute under Contract
No. DE-AC05-76RL01830,
and by the U.S.~National Science Foundation under award No.~NSF-PHY2111213.
\end{acknowledgments}

\appendix


\bibliography{sapphire_naa_tunl}

\end{document}